\documentclass[11pt]{scrartcl}

\usepackage[utf8]{inputenc}
\usepackage{upgreek}
\usepackage{amsmath}

\usepackage{cite}

\usepackage{nameref,hyperref}

\usepackage{color}

\usepackage{graphicx}

\usepackage{wrapfig}

\begin{document}
\vspace{.7cm}
\noindent\huge{\textbf{Demonstration of the length stability\break requirements for ALPS~II with a\break high finesse 9.2\,m cavity}}\\

\noindent\large\textbf{Jan H. P\~old,\textsuperscript{1,2*} and Aaron D. Spector\textsuperscript{3}}\vspace{-.2cm}\\

\noindent\normalsize\textsuperscript{1}Max Planck Institute for Gravitational Physics (Albert Einstein Institute), Callinstra\ss e~38, D-30167 Hannover, Germany\\
\noindent\normalsize\textsuperscript{2}Leibniz Universit\"{a}t Hannover, Welfengarten~1, D-30167 Hannover, Germany\\
\noindent\normalsize\textsuperscript{3}Deutsches Elektronen-Synchrotron (DESY), Notkestra\ss e~85, D-22607 Hamburg, Germany\\
\noindent\textsuperscript{*}jan.poeld@aei.mpg.de\vspace{.7cm}\\


\begin{minipage}{.95\textwidth}
Light-shining-through-a-wall experiments represent a new experimental approach in the search for undiscovered elementary particles not accessible with accelerator based experiments. The next generation of these experiments, such as ALPS~II, require high finesse, long baseline optical cavities with fast length control. In this paper we report on a length stabilization control loop used to keep a 9.2\,m cavity resonant. The finesse of this cavity was measured to be 101,300$\pm$500 for 1064\,nm light. Fluctuations in the differential cavity length as seen with 1064\,nm and 532\,nm light were measured. Such fluctuations are of high relevance, since 532\,nm light will be used to sense the length of the ALPS~II regeneration cavity. Limiting noise sources and different control strategies are discussed, in order to fulfill the length stability requirements for ALPS~II.
\end{minipage}

\vspace{.7cm}
%%%%%%%%%%%%%%%%%%%%%%%%%%  body  %%%%%%%%%%%%%%%%%%%%%%%%%%
\normalsize
\section{Introduction}
Axion-like particles~\cite{alps} represent an extension to the standard model of particle physics that could explain a number of astrophysical phenomena including the transparency of the universe for highly energetic photons~\cite{TEV} as well as excesses in stellar cooling~\cite{stellarcool}. These particles are characterized by their low mass, $m<1$\,meV, and weak coupling to two photons, $g<10^{-10}$\,GeV$^{-1}$. The most prominent axion-like particle is the axion itself which is predicted to preserve the so called charge-parity conservation of Quantum chromodynamics~\cite{axion}. Axions and axion-like particles are also excellent candidates to explain the dark matter in our universe~\cite{DM}. 

Light-shining-through-a-wall experiments attempt to measure the interaction between axion-like particles and photons by shining a laser through a strong magnetic field at an optical barrier. This will generate a flux of axion-like particles traveling through the optical barrier to another region of strong magnetic field on the other side of the barrier. Here, some of the axion-like particles will reconvert to photons that can be measured.

Any Light Particle Search (ALPS)~II~\cite{alpstdr} is a light-shining-through-a-wall experiment that is currently being set up at DESY in Hamburg. It uses strong, superconducting dipole magnets and a high power laser with 122\,m cavities on either side of the optical barrier to boost the conversion probability of photons to axion-like particles and vice versa. The cavity before the barrier is called the Production Cavity (PC), while the cavity after the barrier is called the Regeneration Cavity (RC). 

In order for ALPS~II to reach a sensitivity necessary to probe the photon couplings predicted by the aforementioned astrophysical phenomena the experiment must employ long baseline, high finesse cavities. This is because increasing the number of photons in the PC increases the axion-like particle flux, while the finesse of the RC amplifies the probability that axion-like particles will reconvert to photons~\cite{rc}. A demonstration of the optical subsystems for ALPS~II is currently taking place in a 20\,m test facility, referred to as ALPS~IIa~\cite{20mcavity}, whereas the 245\,m full-scale experiment will be called ALPS~IIc.

In the current ALPS~IIc design, the PC will be seeded with 30\,W generated from a high power laser operating at 1064\,nm~\cite{Frede:07}. The cavities will be stabilized using  the Pound-Drever-Hall (PDH) technique~\cite{Drever,Black}. With a power buildup factor of 5,000 the PC will achieve a nominal circulating power of 150\,kW. For the resonant enhancement of the reconversion process it is crucial that the light circulating inside the PC is simultaneously resonant in the RC. Active stabilization systems will be required to suppress the differential length noise between the cavities and maintain the dual resonance condition. Two detection methods with very different systematics are planed for ALPS~II. A heterodyne detection scheme \cite{zachhet} and a transition edge sensor \cite{jandetes} will be installed subsequently. For the transition edge sensor the length sensing of the RC cannot use 1064\,nm light to generate an error signal for the feedback control loop as this would be indistinguishable from the regenerated light. Instead 1064\,nm light that is offset phase locked to the light transmitted by the PC will be frequency doubled in front of the optical barrier and the length stabilization system will utilize 532\,nm light. According to the ALPS~IIc design, the length stabilization system must ensure that the power buildup for the regenerated photons stays within 90\% of its value on resonance. Even though a seismically quiet environment is chosen for the ALPS~II experiments, this sets challenging requirements on the bandwidth of the length control loop and requires a custom made, piezo controlled length actuator.

The length stability requirement calls for a differential length noise with an RMS value of less than 0.6\,pm between the PC and the RC. The ALPS~II RC will have a finesse of  $\sim$120,000 for 1064\,nm light and a linewidth of 10\,Hz. Circulating fields in each of the cavities will propagate through 560\,Tm of magnetic field length. Considering all of the parameters given above ALPS~IIc will achieve a sensitivity of g$_{\alpha\gamma\gamma}$=$2\times10^{-11}$\,GeV$^{-1}$ for the coupling constant of photons to axion-like particles with masses up to 0.1\,meV. While a detailed overview and status report on ALPS~II is given in~\cite{alpstdr} and~\cite{patras18} this paper focuses on the implementation and characterization of the length stabilization system of the ALPS~IIa RC.

\section{Setup}
\label{sec:setup}
The ALPS~IIa RC is being characterized with two figures of merit: finesse for 1064\,nm light and differential length noise. For the characterization of the differential length noise a high bandwidth control loop with 532\,nm light stabilizes the length of the RC. The error point noise of this setup can be calibrated to provide an in-loop measurement of the suppressed length noise of the cavity. Furthermore, locking a separate 1064\,nm laser to the cavity revealed noise sources that were not observable with the in-loop measurement. The finesse of the RC for 1064\,nm light is characterized by measuring the cavity storage time.

\begin{figure}[!t]
\centering\includegraphics[width=\textwidth]{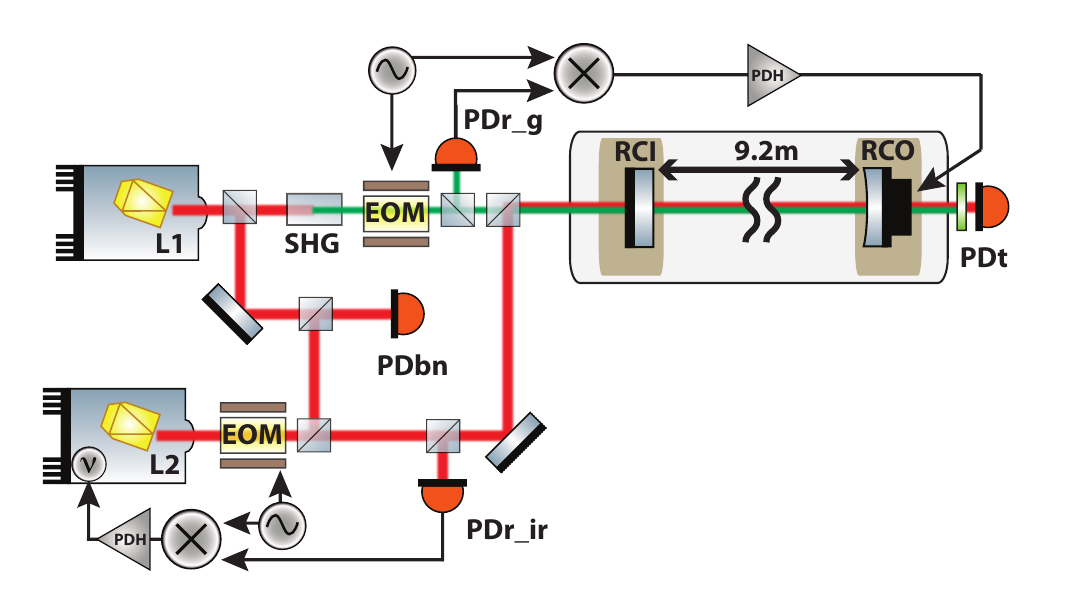}
\caption{\textbf{Experimental setup.} \textit{Length control of the RC:} The cavity consists of the mirrors RCI and RCO as well as a PDH feedback control loop. A high bandwidth length actuator is attached to RCO. The laser beam from laser L1 is frequency doubled in a temperature controlled SHG. An EOM imprints phase modulation sidebands on the laser beam and the photodetector PDr\_g is used to sense the PDH error signal. \textit{Laser frequency feedback:} For the high finesse cavity operation with 1064\,nm light the frequency of laser L2 follows the RC. The feedback control loop uses PDr\_ir as a sensor and photodetector PDt in transmission of the cavity is used to measure the storage time. \textit{Beatnote frequency measurement:} The beatnote signal between L1 and L2 is measured with a high bandwidth photodetector (PDbn).}
\label{fig:setup}
\end{figure}

A 500\,mW non-planar-ring-oscillator L1 at a wavelength of 1064\,nm is used to implement the length lock of the RC. It seeds a periodically poled potassium titanyl phosphate crystal which generates 100\,$\mu$W of 532\,nm light in a single-pass second harmonic generation (SHG) (see schematic in Fig. \ref{fig:setup}). An electro-optic modulator (EOM) adds phase modulation sidebands before the light enters the optical cavity. 

The two cavity mirrors are mounted on separate optical tables 9.2\,m apart from each other and within a common vacuum system. A rubber material in the feet of the optical table provides  dampening above 100\,Hz. For the measurements the system was pumped down to $1\times10^{-5}$\,mbar in order to minimize acoustic couplings. The entire experiment is located in a clean and temperature controlled environment which is similar to the conditions we anticipate for the ALPS~IIc experiment.

The cavity input mirror RCI is flat while the cavity end mirror RCO has a radius of curvature of 19.7$\pm$0.1\,m. This configuration yields a beam radius on RCI of 1.82$\pm$0.01\,mm and on RCO of 2.51$\pm$0.01\,mm for 1064\,nm light, respectively. Each mirror has a diameter of 50.8\,mm with a mass of 43\,g and features a dichroic coating. The mirror size was chosen to avoid diffraction losses in ALPS~IIc. RCI has a nominal power transmission of 25\,ppm for 1064\,nm and 5\,\% for 532\,nm light. The RCO coating has a power transmission of 3\,ppm for 1064\,nm and 1\,\% for 532\,nm light. The free spectral range is 16.2\,MHz. 

A second laser L2 (see Fig. \ref{fig:setup}) seeds the cavity with infrared light. Photodetectors PDr\_g and PDr\_ir sense the beat signal between the directly reflected field of the cavity and a fraction of the circulating field that is transmitted through RCI for green and infrared light, respectively. Each signal at the output of the photodetector is demodulated, amplified in the PDH servo electronics and sent to the actuator. PDbn senses the beatnote signal of L1 and L2. In addition, photodetector PDt monitors the power in transmission of the cavity and is also used to perform a measurement of the storage time.

\section{High finesse cavity characterization}
\label{sec:fc}
State-of-the-art optics with ultra low losses are required to construct a cavity with a finesse of  $\sim$120,000 for the ALPS~II RC~\cite{alpstdr}. These types of cavities must be set up in vacuum to avoid any kind of dust particles contaminating the mirror surfaces and avoid scattering of the intra-cavity light.

\begin{figure}[!t]
\centering\includegraphics[width=\textwidth]{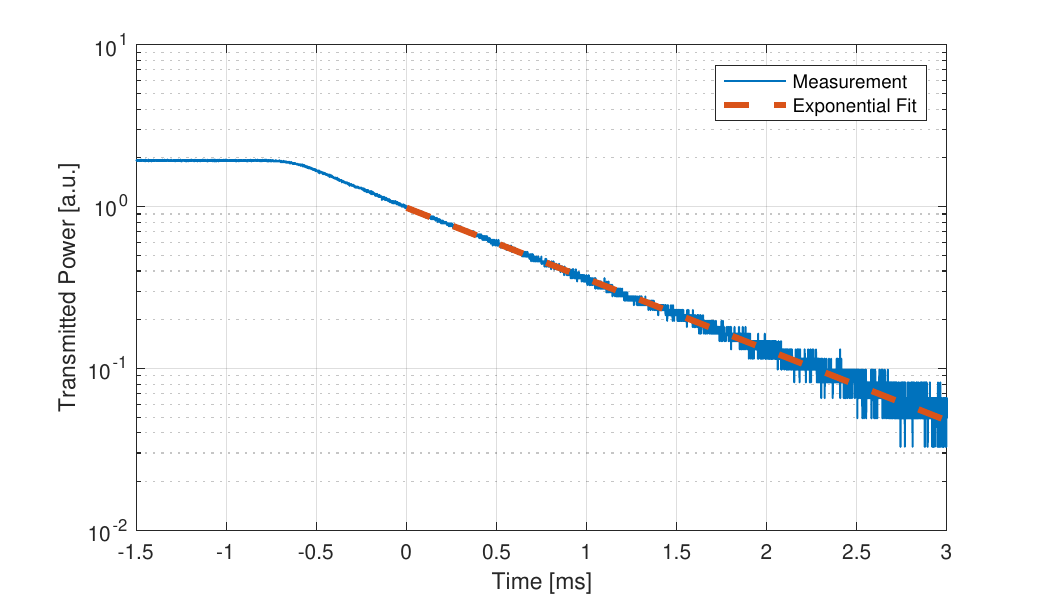}
\caption{\textbf{Storage time measurement.} The cavity storage time is a measurement of the exponential decay of the transmitted power when the laser shutter is closed.}
\label{fig:storage}
\end{figure}

Once the laser is frequency locked to the cavity the input light is blocked by suddenly closing the laser shutter. Then the exponential decay of the transmitted power is measured to determine the cavity storage time. The following function was fit to the data~\cite{cavloss}:
\begin{equation}
P_{\text{trans}}(t)=P_0 g^2 T_{\text{in}} T_{\text{out}}\text{exp}\left( -\frac{2t}{\tau_{\text{storage}}}\right)
\end{equation}
\noindent In this equation $g^2$ is the cavity gain factor, $P_0$ is the initial power, $T_{\text{in}}$ and $T_{\text{out}}$ are the power reflectivities of the input and output mirror, respectively.

Figure \ref{fig:storage} shows the result of one of the storage time measurements. An average of ten measurements yielded a storage time $\tau_{\text{storage}}$ of 1.99$\pm$0.01\,ms. The fit considers data points when the power in the cavity dropped by a factor of two since it takes some time until the shutter has blocked the entire input beam. Applying equations from reference~\cite{cavloss} yields a finesse of 101,300$\pm$500 and the roundtrip losses are 33$\pm$1\,ppm. This does not include the transmissivities of the mirrors. We believe that most of the losses are due to scattering caused by low spatial frequency surface roughness of the mirrors. While the result strongly depends on the beam spot position on the mirrors, the measurements were performed at a position that gave the longest storage time.

\section{High bandwidth cavity lock}
\label{sec:lc}
One of the key parameter for the ALPS~II sensitivity is the differential length stability between the PC and the RC. The differential RMS length noise between these two cavities must be suppressed to less than 0.6\,pm. This will ensure that light that is resonant with the PC will experience 90\,\% of the maximum power buildup in the RC. This is what we refer to as the dual resonance condition. As mentioned earlier the PDH error signal for the RC is generated using 532\,nm light. 

Based on the transmission values of the cavity mirrors for 532\,nm light the finesse is 102 and the linewidth 158.6\,kHz in ALPS~IIa. The low finesse for 532\,nm light was chosen such that only a minimum amount of light is circulating in the RC. A conditionally stable control loop design with two integrators is used to suppress the noise as much as possible. In order to smoothen the transfer function of the piezo actuator attached to RCO and have less impact from the piezo resonances a digital filter was inserted into the control loop. The filter coefficients were chosen such that they inverted the piezo transfer function. Consequently, this optimized the phase and gain margin of the control loop. A unity-gain-frequency of 4\,kHz was achieved with a phase margin of 20\,deg.

\begin{figure}[!t]
\centering\includegraphics[width=\textwidth]{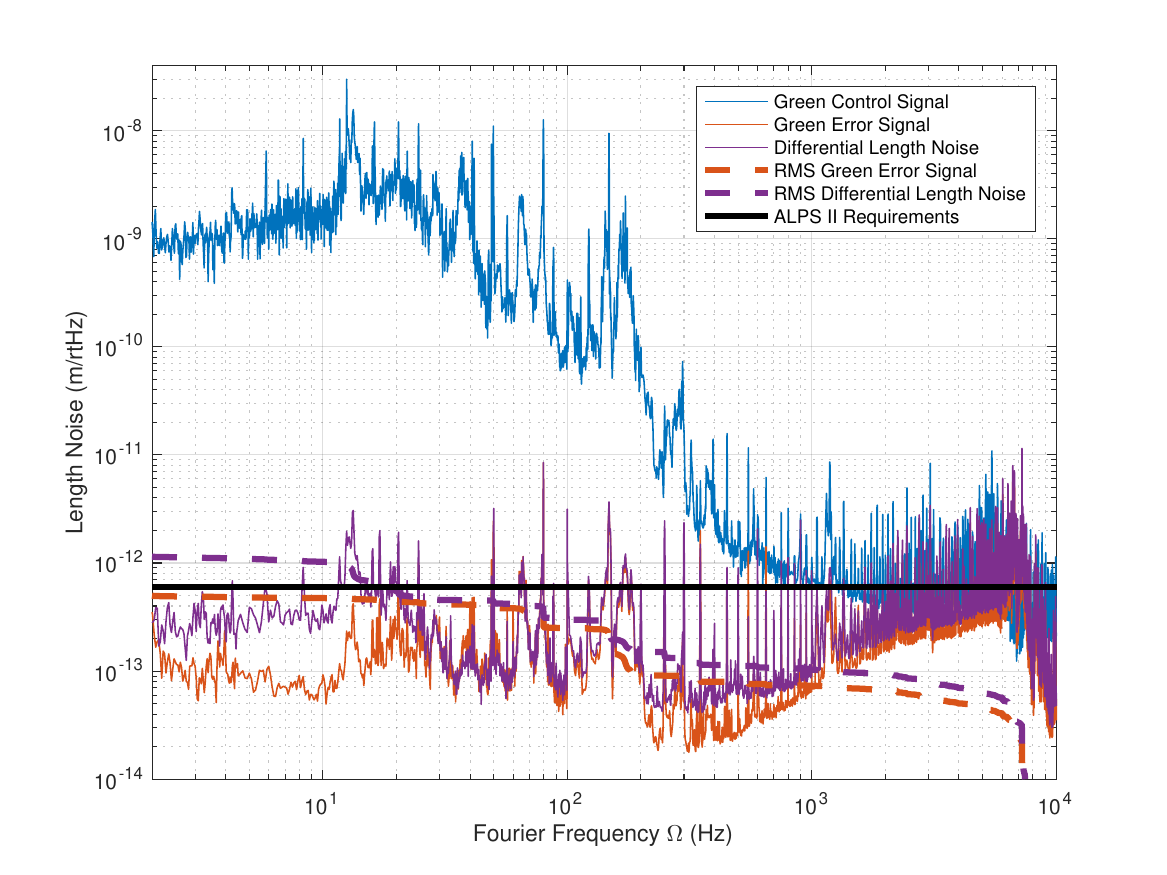}
\caption{\textbf{Length noise measurement.} Amplitude spectral densities of control and error signal of the PDH control loop calibrated in length noise as well as the beatnote frequency measurement representing the differential length noise. For a comparison with the ALPS~II requirements the error signal and the beatnote frequency measurement are filtered by a cavity pole frequency of 6\,Hz in post processing and its corresponding integrated RMS is shown with the dotted lines.}
\label{fig:fqnoise}
\end{figure}

The length actuator is a piezo ceramic (Physik Instrumente GmbH \& Co. KG). We designed a custom mount to hold a stack consisting of the piezo, the cavity end mirror RCO and a wave washer. The stack is kept in place by exerting pressure on the wave washer with a retaining ring that is screwed into the mount. This also has the effect of preloading the piezo. The force exerted on the stack was optimized such that the resonances of the system were pushed as high as possible. It was also important not to over tighten the retaining ring as this reduced the performance of the length acutator. The result for the optimized setup contains the first resonance at 4.9\,kHz.

\subsection*{In-loop measurement}
Figure \ref{fig:fqnoise} shows a spectral density of the control and error signal displayed in terms of length noise of the cavity. As already mentioned in \cite{20mcavity} the control signal is dominated by seismic noise up to 1\,kHz and by laser frequency noise above 1\,kHz. The error signal represents an in-loop measurement of the suppressed length noise. Electronics noise from the digital controller effected the measurement below 10\,Hz. This will be addressed by using a different digital control system, however this noise still does not prevent the in-loop measurements from meeting the requirements.

Cavities exhibit a passive low pass filter property for their circulating fields. Hence, the frequency noise of the input field is suppressed at Fourier frequencies above the cavity pole~\cite{Mueller}. In order to predict the impact to ALPS~IIc the error signal noise is therefore filtered in post processing by the expected filter property of the ALPS~IIc RC. This consists of a low pass with a pole frequency of 6\,Hz, assuming a Finesse of 100,000 as reported on in the previous section. The RMS projection shows that the control loop has sufficient gain to meet the length noise requirements for ALPS~IIc considering similar uncontrolled length noise conditions as in the ALPS~IIa lab~\cite{dome}.

\subsection*{Beatnote frequency measurement}
\label{sec:bfm}
Since the measurement in the previous section was an in-loop measurement it was important to confirm the result with an out-of-loop measurement. This was performed with a second 1064\,nm laser (L2). While the cavity length is locked to the frequency doubled laser L1, the frequency of the second laser L2 is locked to the cavity in order to simulate the light that comes from the PC and is phase locked to L1. The $\sim$50\,MHz infrared beatnote frequency is monitored with a fast photodetector PDbn and demodulated down to 100\,kHz by mixing it with a stable reference. A time series of the 100\,kHz signal is then recorded and its frequency noise is analyzed in post-processing.

\begin{figure}[!t]
\centering\includegraphics[width=\textwidth]{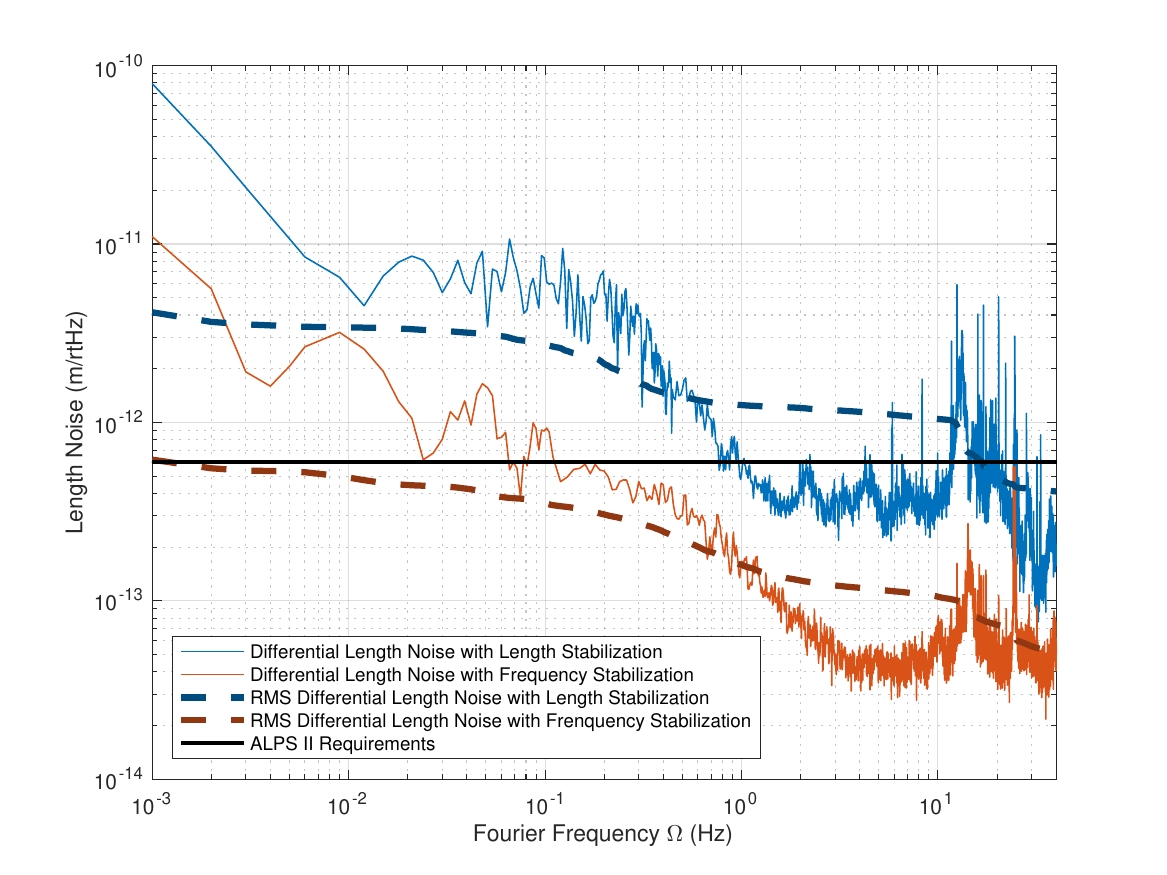}
\caption{\textbf{Out-of-loop measurement.} Beatnote frequency measurement for the length and frequency feedback and corresponding RMS filtered by the infrared RC cavity pole for ALPS~IIc.}
\label{fig:ool}
\end{figure}

The measurement is displayed in Fig.~\ref{fig:fqnoise}. Unexplained out-of-loop noise enters below 25\,Hz and above 200\,Hz. The filtered RMS noise, displayed in the corresponding dotted line, exceeds the ALPS~II length noise requirements by a factor of roughly two. In order to address the out-of-loop noise noise below 25\,Hz the control signal of the length control loop was fed back to the laser frequency of L1 instead. Thus the bandwidth of the loop could be significantly increased to 40\,kHz. Electronics noise, which limited the in-loop measurement for the length lock below 10\,Hz was substantially lower as the digital controller was not required for this type of control loop.  

Figure \ref{fig:ool} shows the data below 40\,Hz for the length and frequency feedback, respectively. The measurement was loop gain limited above 40\,Hz. While the length stabilization crosses the requirements at 17\,Hz and increases further to an RMS value of 3.5\,pm at 1\,mHz, the frequency stabilization RMS meets the requirements down to 1.3\,mHz. 

It is apparent that the actuation on the piezo increases the out-of-loop noise. In both the frequency and length feedback we believe this noise is related to alignment noise of the cavity coupling to differential length noise between green and infrared eigenmodes. This could be due to the surface roughness of the mirrors. Since the green and infrared eigenmodes have different beam sizes in the cavity they also have different scattering matrix coefficients~\cite{scatter}. Furthermore, as the scattering matrix coefficients are spatially dependent, any movement of the eigenmode along the surface of the mirror will lead to a relative change in the reflected phase for green and infrared even though the eigenmodes themselves remain aligned with respect to each other to first order. While the alignment noise may increase in ALPS~IIc due to the longer baseline, the surface quality of the mirrors is expected to be better. For ALPS~IIc the requirement on the mirror surface roughness is less than two angstrom RMS across the mirror surface. With better mirrors we anticipate that the out-of-loop noise will decrease.

\section{Conclusion}
\label{sec:conc}

For the first time the length stability requirements for ALPS~II have been demonstrated. The control loop actuating on the length of the cavity was capable of maintaining the resonance conditions to an accuracy that is below the ALPS~II requirements by achieving a unity-gain-frequency of 4\,kHz. A customized, high bandwidth length actuator that moves a 50.8\,mm mirror was an essential component of this system. The discovery of the additional out-of-loop noise indicates that it might be necessary to change the control concept for ALPS~II such that the PC length will be actuated on. The PC length sensing will be done with infrared light which avoids differential effects for the green and infrared eigenmodes.

Furthermore, an out-of-loop measurement of the differential length noise of the RC with a feedback to the laser frequency led to a successful demonstration of the ALPS~IIc length noise requirements in the ALPS~IIa environment. If the out-of-loop noise is not reduced in the future, it is an option to open a shutter in the light tight wall roughly every 1000\,s to ensure that the resonance condition for infrared light is still met. This will allow us to set up ALPS~IIc without a dedicated seismic isolation system for the cavity mirrors, as measurements indicate that the seismic noise environment for ALPS~IIc is similar to ALPS~IIa~\cite{dome}.

In addition, the finesse of the ALPS~IIa RC was measured to be 101,300$\pm$500 with a storage time of 1.99$\pm$0.01\,ms. These results are comparable to experiments that employ long baseline, high finesse optical cavities such as gravitational wave detectors~\cite{aligo,sato99}, filter cavities for non-classical light~\cite{filtercav} and vacuum magnetic birefringence experiments~\cite{pvlas}.

These results represent a major milestone for ALPS~II from the previous work~\cite{20mcavity}. The next steps will be towards the identification of the out-of-loop noise sources and their mitigation.

\section*{Funding}
Deutsche Forschungsgemeinschaft (DFG) (SFB 676)\\
Volkswagen Stiftung

\section*{Acknowledgments}
The authors would like to thank the other members of the ALPS collaboration for valuable discussions and support, especially Axel Lindner and Benno Willke. This work would not have been possible without the wealth of expertise and hands-on support of the technical infrastructure groups at DESY.

%%%%%%%%%%%%%%%%%%%%%%% References %%%%%%%%%%%%%%%%%%%%%%%%%


\begin{thebibliography}{99}
\small
\bibitem{alps} C. Patrignani et al (Particle Data Group). ``Review of Particle Physics", Chin. Phys. C, 2016, {\textbf {40}} (10): 100001
\bibitem{TEV}M. Meyer, D. Horns and M. Raue, ``First lower limits on the photon-axion-like particle coupling from very high energy gamma-ray observations", Phys. Rev. D, {\textbf 5}, 035027, (2013).
\bibitem{stellarcool}M. Giannotti, I. Irastorza, J. Redondo and A. Ringwald,  ``Cool WISPs for stellar cooling excesses", J. Cosmol. Astrop. Phys.,  {\textbf 5}, 57 (2016).
\bibitem{axion} R. D. Peccei and H. R. Quinn, ``CP Conservation in the Presence of Pseudoparticles" Phys. Rev. Lett. {\textbf {38}}, 1440 (1977). 
\bibitem{DM}L.F. Abbott and P. Sikivie, ``A cosmological bound on the invisible axion" Phys. Lett. B {\textbf{1}}, 133--136 (1983).
\bibitem{alpstdr} R. B\"ahre, B. D\"obrich, J. Dreyling-Eschweiler, S. Ghazaryan, R. Hodajerdi, D. Horns, F. Januschek, E. -A. Knabbe, A. Lindner, D. Notz, A. Ringwald, J. E. von Seggern, R. Stromhagen, D. Trines and B. Willke, ``Any light particle search II - Technical Design Report," J. Inst. {\textbf 8} (9),T09001 (2013).
\bibitem{rc}F. Hoogeveen and T. Ziegenhagen, ``Production and detection of light bosons using optical
resonators," Nucl.Phys. B 358, 3--26 (1991).
\bibitem{20mcavity}A. D. Spector, J. H. P\~old, Robin B\"ahre, A. Lindner, and B. Willke, ``Characterization of optical systems for the ALPS~II experiment," Opt. Express 24, 29237--29245 (2016).
\bibitem{Frede:07} M. Frede, B. Schulz, R. Wilhelm, P. Kwee, F. Seifert, B. Willke and D. Kracht, ``Fundamental mode, single-frequency laser amplifier for gravitational wave detectors," Opt. Express 15 {\textbf 2}, 459--465 (2007).
\bibitem{Drever}R. W. P. Drever, J. L. Hall, F. V. Kowalski, J. Hough, G.M. Ford, A. J. Munley and H. Ward,  ``Laser phase and frequency stabilization using an optical resonator," Appl. Phys. B 31 {\textbf 2}, 97--105 (1983).
\bibitem{Black}E. D. Black, ``An introduction to Pound-Drever-Hall laser frequency stabilization," Am. J. Phys {\textbf {69}}, 79 (2001).
\bibitem{zachhet}Z.~Bush, S.~Barke, H.~Hollis, A.~D.~Spector, A.~Hallal, G.~Messineo, D.~B.~Tanner, and G.~Mueller, ``Coherent detection of ultraweak electromagnetic fields," Phys. Rev. D {\textbf {99}}, 022001 (2019).
\bibitem{jandetes}J.~Dreyling-Eschweiler, N.~Bastidon, B.~D\"{o}brich, D.~Horns, F.~Januschek, and A.~Lindner, ``Characterization, 1064 nm photon signals and background events of a tungsten TES detector for the ALPS experiment, J. Mod. Opt., {\textbf {62}} (14), 1132--1140 (2015).
\bibitem{patras18} A. D. Spector for the ALPS collaboration, ``ALPS~II status report," arXiv:1906.09011 (2019).
\bibitem{cavloss}T. Isogai, J. Miller, P. Kwee, L. Barsotti and M. Evans, ``Loss in long-storage-time optical cavities," Opt. Express 21, 30114--30125 (2013).
\bibitem{Mueller}C.~L. Mueller, M.~A. Arain, G.~Ciani, R.~T. DeRosa, A.~Effler, D.~Feldbaum, V.~V. Frolov, P.~Fulda, J.~Gleason, M.~Heintze, K.~Kawabe, E.~J. King, K.~Kokeyama, W.~Z. Korth, R.~M. Martin, A.~Mullavey, J.~Peold, V.~Quetschke, D.~H. Reitze, D.~B. Tanner, C.~Vorvick, L.~F. Williams, and G.~Mueller, ``The advanced {LIGO} input optics," Rev. Sci. Instrum. \textbf{87}, 014502 (2016).
\bibitem{dome}D.~Miller, ``Seismic noise analysis and isolation exemplary shown for the ALPS experiment at DESY," PhD thesis, Leibniz Universit\"{a}t Hannover (2019).
\bibitem{scatter}R.~Mavaddat, ``Scattering matrix analysis of optical resonators with cavity imperfections," Opt. Commun., {\textbf {137}} (4), 367--381 (1997).
\bibitem{aligo}The LIGO Scientific Collaboration, ``Advanced LIGO," Classical Quant. Grav. \textbf {32}, 074001 (2015).
\bibitem{sato99}S. Sato, S. Miyoki, M. Ohashi, M. Fujimoto, T. Yamazaki, M. Fukushima, A. Ueda, K. Ueda, K. Watanabe, K. Nakamura, K. Etoh, N. Kitajima, K. Ito, and I. Kataoka, ``Loss factors of mirrors for a gravitational wave antenna," Appl. Opt. 38, 2880-2885 (1999).
\bibitem{filtercav}M. Evans, L. Barsotti, P. Kwee, J. Harms and H. Miao, ``Realistic filter cavities for advanced gravitational wave detectors," Phys. Rev. D 88 {\textbf 2}, 022002 (2013).
\bibitem{pvlas}F. Della Valle, E. Milotti, A. Ejlli, U. Gastaldi, G. Messineo, L. Piemontese, G. Zavattini, R. Pengo, and G. Ruoso,``Extremely long decay time optical cavity," Opt. Express 22, 11570-11577 (2014).
\end{thebibliography}
\end{document}